      [1999/12/01 v1.4c Il Nuovo Cimento]
\documentclass{cimento}

\usepackage{graphicx}
\usepackage{amssymb}
\usepackage{amsmath}

\title{Comments on theory of volume reflection and
radiation in bent crystals}
\author{M.~V.~Bondarenco } \instlist{\inst{} Kharkov Institute of Physics \& Technology,
1 Academic Str., Kharkov 61108, Ukraine }
\PACSes{\PACSit{61.85.+p}{bent crystal; volume reflection; coherent
bremsstrahlung}}
\begin{document}

\maketitle

\begin{abstract}
Recent theoretical results on charged particle interaction with
planarly oriented thin bent crystals are reviewed, with the emphasis
on dynamics in the continuous potential. Influence of boundary
conditions on the volume-reflected beam profile is discussed. Basic
properties of coherent bremsstrahlung in a bent crystal are
highlighted.
\end{abstract}

\section{Introduction}

Passage of charged particles through a bent crystal, even in a
planar orientation, is a complex phenomenon, theoretical treatment
of which used to rely on computer simulation. However, sometimes
complexity begets simplicity, which may permit analytic advances --
like it was commonly practiced before the advent of computers.
Actually, computer and analytic calculations complement one another:
analytic formulas provide general understanding of the phenomenon,
and may serve for experiment planning, while computer can offer
precise predictions for an established experimental setting.

An example of a problem where analytic approach did bring fruit is
fast charged particle passage through a planarly oriented uniformly
bent crystal. Even though the planar orientation and uniformity of
the bending enormously simplify the dynamics, reducing it to radial
1d, yet there are impediments for solution of the whole beam-crystal
interaction problem: how to analytically describe particle passage
through many inter-planar intervals, and how to analytically average
over the initial particle impact parameters. Alleviation came from
the use of a simplified (parabolic) model for inter-planar
continuous potential, which made analytic solution for single volume
reflection (VR) feasible \cite{ref:Bond-VR}. It appears that precise
inter-planar potential shape is of minor consequence anyway, and the
piecewise harmonic potential model yields fair agreement with the
experiment.

A different story is the bremsstrahlung emitted by the fast particle
in a bent crystal. There, a large contribution comes from the dipole
coherent bremsstrahlung, which can be described for an arbitrary
inter-planar potential. Analytic theory of dipole coherent radiation
in bent crystals (CBBC) was constructed in \cite{ref:Bond-CBBC}.

The present short note surveys the physical picture of volume
reflection and radiation in a bent crystal, in the pure continuous
potential.

\section{Volume reflection}

\subsection{Essence of the effect}
VR effect manifests itself in the same region of $E/R$ ratio as
channeling, with the proviso that the particle entrance angle with
respect to active crystallographic planes must be well above
critical. However, the correlation between the particle deflection
angle and the crystal curvature for VR appears to be more intricate
than that for channeling:
\begin{enumerate}
  \item in VR particles deflect to the side opposite to that
of the crystal bending;
  \item the
deflection angle tends to a finite limit with the increase of
$R/R_c$ ratio, although $R=\infty$ corresponds to a straight crystal
and no net deflection; this limiting value is of the order of
critical angle $\theta_c$ both for positive and for negative
particles, though with different numerical coefficients.
\end{enumerate}

\begin{figure}[h]
\begin{minipage}{18pc}
\includegraphics{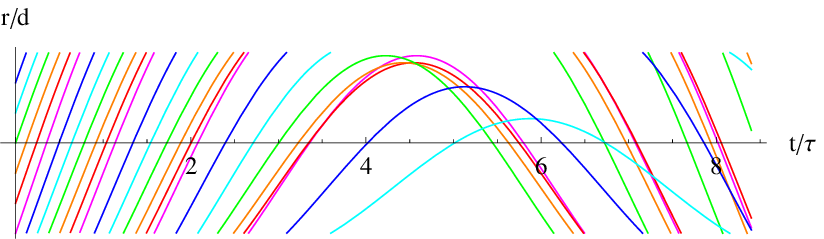}\\
\includegraphics{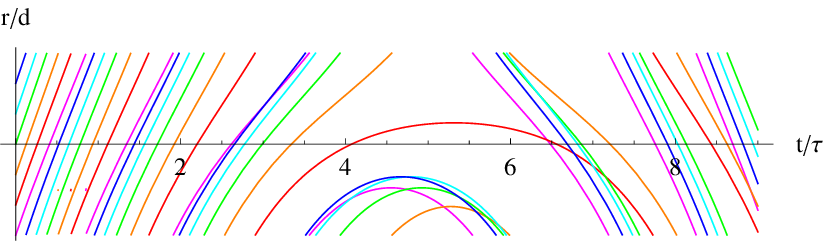}
\end{minipage}\hspace{2pc}%
\begin{minipage}{14pc}
\includegraphics[width=12pc]{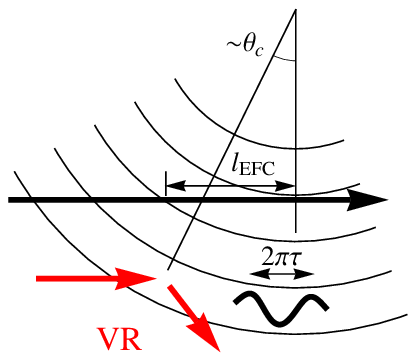}
\end{minipage}
\caption{\label{fig:Vol-vs-Edge} Left-side frame -- trajectory
filling of an exemplary inter-planar interval by an initially
transversely uniform particle beam (top -- positive particles,
bottom -- negative particles). The deficit of positively directed
force in VR area causes the negative net deflection of the beam.
Right-side frame -- comparison of ultra-high energy ($R\ll 4R_c$,
$l_{\mathrm{EFC}}\ll 2\pi\tau$) with moderately high energy ($R\gg
4R_c$, $2\pi\tau\ll l_{\mathrm{EFC}}$) particle passage. In the
latter case the particle net deflection angle is non-zero and
$\sim\theta_c$, regardless of the particle charge sign.}
\end{figure}

The qualitative explanation of the opposite deflection direction is
given in \cite{ref:Scandale-Rdependence}: ``For a small crystal
curvature, that is $R\gg R_c$, the turning points of all particles
gather in a narrow region near the inner wall of a planar channel.
The strong electric field of the crystal plane is directed along $R$
and at the turning points it imparts to the particle an angular
deflection towards the opposite direction with respect to the
crystal bending, producing volume reflection." For explanation of
feature 2 the latter argument is not sufficient yet. It is known
that for negative particles the force in the reflection point is
much smaller than for positive ones, but nevertheless, the
deflection angle for these cases is of the same order. For a closer
look, let us draw a graph of the family of particle trajectories in
an exemplary inter-planar interval (Fig.~\ref{fig:Vol-vs-Edge}). The
figure reveals that in inter-planar channels there emerge regions
devoid of particles. This entails deficit of force of a definite
(positive) sign acting on the beam as a whole; hence, the mean
deflection angle must be negative. At sufficiently large $R$, the
void transverse dimension is $\sim d$, so $F\sim F_{\max}$, and the
longitudinal dimension is $\sim\tau=\sqrt{R_cd/2}$ (the channeling
oscillation timescale)\footnote{Note that this scale is $R/R_c$
times shorter than scale $R\theta_c$ within which the particle
trajectory substantially differs from a straight line. At $R\gg
R_c$, in principle one should mind the existence of two longitudinal
scales in the volume reflection phenomenon.}; none of these
parameters involve $R$, therefore, there is a limiting value
$\lim_{R\to\infty}\left\langle\theta\right\rangle_b$ independent of
$R$:
$\left\langle\theta\right\rangle_b\sim\frac1EF_{\max}\tau=\theta_c$.

Of course, if one overbends the crystal, all the intra-crystal space
becomes uniformly covered by the particle flow, then voids and
therewith the VR effect disappear. In
\cite{ref:Bond-VR,ref:Bond-CBBC} it was proved that if $E\to\infty$,
i.e., $R/R_c\to0$, then $\int db \theta(b)= 0$.

\emph{Ex adverso} the existence of VR effect can  be explained more
concisely: When a high-energy particle passes through a bent
crystal, after its entrance the angle between its velocity and the
active planes decreases. When this angle becomes $\sim\theta_c$, the
particle experiences strong influence of the continuous potential
field on the length $\tau$. If this length is shorter than
$l_{\mathrm{EFC}}=\sqrt{2Rd}$, it would be capable of making the
particle channeled. But for motion in a stationary potential, an
unbound particle cannot pass into bound state. Hence, at
$\pi\tau<l_{\mathrm{EFC}}$, i.e. $R>\frac{\pi^2}4R_c$, the particle
net inclination angle can not become smaller than $\theta_c$, which
implies that the particle will reflect when the angle becomes
$\sim\theta_c$ -- see right-side frame of
Fig.~\ref{fig:Vol-vs-Edge}.



\subsection{Boundary condition sensitivity}

Although the term `volume' reflection signifies that the deflection
arises somewhere in the crystal volume, and therefore should not
depend on the crystal boundaries, but to some degree, the boundary
effects still manifest themselves when investigating the final beam
profile detail.

\begin{figure}[h]
\includegraphics[width=30pc]{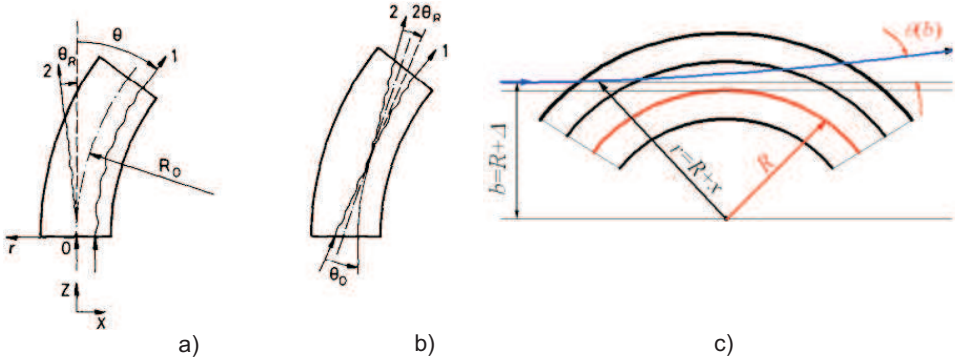}
\caption{\label{fig:bound-a,b}Possible boundary conditions for VR in
a uniformly bent crystal. Left-side frame, cases a) and b) (the
figure is taken from \cite{ref:Tar-Vor}) -- particle entrance
through a flat front face of a thin bent crystal. At modern
practice, the $x$-dimension of the crystal is actually much wider
than its $z$-dimension. Right-side frame (figure taken from
\cite{ref:Shulga-talk}) -- particle entrance through the curved
lateral surface of a longitudinally extended crystal; this type of
boundary conditions is also adopted in \cite{ref:Kovalev}.}
\end{figure}

First of all, even for a uniformly bent crystal, there are several
options for particle beam entrance to the crystal -- see
Fig.~\ref{fig:bound-a,b}. At practice one usually deals with case
b), when the particle entrance angle $\theta_0$ is much greater than
critical; the corresponding theory was worked out in
\cite{ref:Maisheev-VR,ref:Bond-VR}. But at channeling experiments,
there also arises case a), which was studied in the pioneering work
\cite{ref:Tar-Vor}. In the latter case, the particles enter the
crystal tangentially to the active planes, so, at $R\gg R_c$ most of
the particles are channeled, but a small fraction of particles
hitting one-sided vicinity of the top of potential barrier belongs
to volume reflection. For the latter fraction, it is supposed that
owing to symmetry of the trajectory in a central field with respect
to the reflection (minimal-radius) point, the deflection angle
$\theta$ for case a) is half the value for case b). But yet there is
a slight dependence of the deflection angle on the impact parameter,
and it is this dependence which determines the final beam shape. So,
a question remains, whether the final beam profiles for cases a) and
b) are similar.

\begin{figure}[h]
\begin{minipage}{14pc}
\includegraphics[width=14pc]{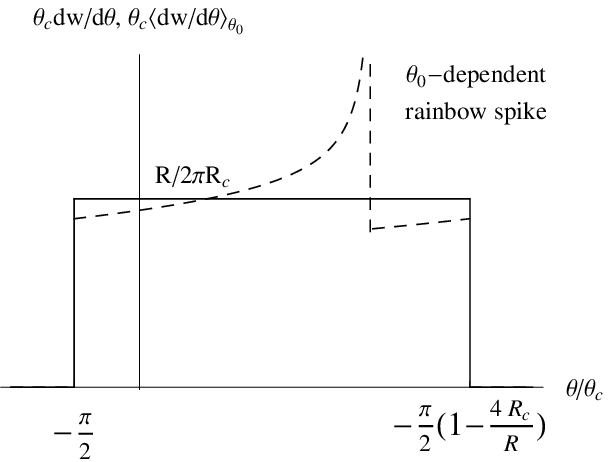}
\end{minipage}\hspace{2pc}%
\begin{minipage}{14pc}
\includegraphics[width=15pc]{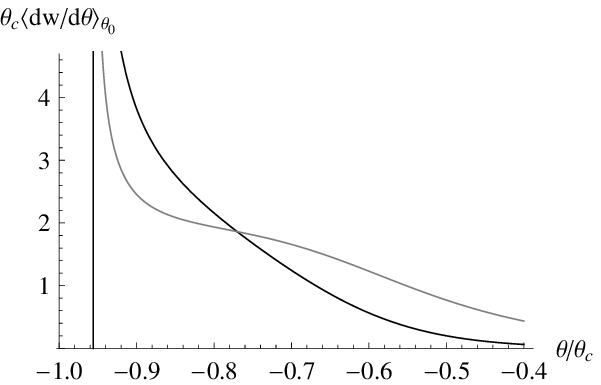}
\end{minipage}
\caption{\label{fig:profile}Left-side frame -- angular distribution
in a VR beam for positive particles, at $R>4R_c$, without account of
multiple scattering. Solid curve -- distribution averaged over a
small variations of initial large angle $\theta_0$, dashed curve --
for a definite $\theta_0$. Right-side frame -- the angular
distribution for negative particles, plotted for $R=25R_c$. Black
curve -- for boundary condition of type b) (averaged over a small
vicinity of large angle $\theta_0$), gray curve -- for case a) and
stretched in $\theta$ by the factor of 2. The peaks on the figures
are due to rainbow phenomena\cite{ref:Bond-VR}.}
\end{figure}

To establish correspondence between the profiles for cases a) and
b), neglecting meanwhile the multiple scattering,
note that $\theta$ depends on the impact parameter $b$ only through
the transverse energy
$E_\perp=E\frac{\theta_0^2}2+V_{\mathrm{eff}}(-b)$, with
$V_{\mathrm{eff}}(-b)=V(-b)+\frac{Eb}{R}$, whereas $\theta(E_\perp)$
is a universal periodic function of $E_{\perp}$ with the period
$\Delta E_{\perp}=\frac{Ed}{R}\sim\frac{R_c}{R}V_0\ll V_0$
\cite{ref:Maisheev-VR}. However, dependencies $E_\perp(b)$ are
different under different boundary conditions. For case b),
$\theta(E_\perp)$ spans many periods as $b$ varies from $-\frac{d}2$
to $\frac{d}2$, and $E_\perp(b)$ dependence, basically, is linear
within each period of $\theta(E_\perp)$. But for case a)
$\theta(E_\perp)$ spans only one period, and the dependence of
$V_{\mathrm{eff}}$ near its top is essential. Now, for positive
particles this behavior is close to linear, too, and so for positive
particles boundary conditions a) and b) yield basically isomorphic
profiles. On the contrary, for negative particles
$V_{\mathrm{eff}}(b)$ behavior near the top is quadratic. Thus, the
relation between the profiles appears to be
$2\theta^{a)}(\sqrt\nu)=\theta^{b)}(\nu)$, where $\nu$ is the
transverse kinetic energy on the entrance to the reflection
inter-planar interval, rescaled to vary from 0 to 1. See
Fig.~\ref{fig:profile}, right-side frame. \footnote{In paper
\cite{ref:Bond-VR}, footnote 14, it is asserted that for case b)
there is no rainbow, but more precisely, the rainbow significantly
attenuates, though in principle remains.}

As for boundary condition of type c) (see Fig.~\ref{fig:bound-a,b}),
which simplifies the theoretical problem by making it entirely
centrally-symmetric, it yields$\frac{dw}{d\theta}$ similar to the
case b), though indicatrix $\theta(b)$ has a somewhat different
double-periodic behavior \cite{ref:Shulga-talk}.

\subsection{Multiple scattering \label{subsec:mult-scat}}
Intrinsic VR angular distributions shown in Fig.~\ref{fig:profile}
have an interesting edgy structure, for negative particle case being
yet asymmetric and containing a significant rainbow. But at
practice, those distinction features will be obscured by multiple
scattering. However, the extent of the VR area
$2\pi\tau=\pi\sqrt{2?R_cd}\sim50\mu$m is much shorter than the
crystal thickness $L\sim1$mm. Thereat, the multiple scattering
effects accumulate mainly away from the VR point (upstream and
downstream of it). This allows one to view the aggregate
distribution as a convolution of successive scattering probabilities
\cite{ref:Maisheev-VR}:
\begin{equation}\label{convol}
    \frac{dw_{\mathrm{real}}}{d\theta}\approx\int^\infty_{-\infty}
    d\theta_1\frac{e^{-(\theta-\theta_1)^2/2\sigma^2_{\mathrm{am}}}}{\sqrt{2\pi}\sigma_{\mathrm{am}}}\frac{dw_{\mathrm{v.r.}}}{d\theta_1}\quad\,
\left(\int^\infty_{-\infty}d\theta\frac{dw_{\mathrm{real}}}{d\theta}=\int^\infty_{-\infty}d\theta\frac{dw_{\mathrm{v.r.}}}{d\theta}=1\right),
\end{equation}
where ${dw_{\mathrm{v.r.}}}/{d\theta}$ is calculated with the
neglect of multiple scattering, and $\sigma^2_{\mathrm{am}}$ is the
mean square angle of multiple scattering in an amorphous target of
the same thickness. Consideration beyond approximation
(\ref{convol}) will be given elsewhere.

In principle, Eq.~(\ref{convol}) can be inverted; it is noteworthy
that Fourier transforms of ${dw_{\mathrm{real}}}/{d\theta}$ and
${dw_{\mathrm{v.r.}}}/{d\theta}$ are proportional, and should both
have negative regions. But unfortunately, under conditions of binned
measurement of ${dw_{\mathrm{real}}}/{d\theta}$, only first few
moments of the angular distribution can be accessed reliably.


\subsection{Moments of the final angular distribution}

The complete analytic calculation of
${dw_{\mathrm{v.r.}}}/{d\theta}$ was made within the model of
\cite{ref:Bond-VR}. According to this solution, the limiting value
for $\theta_{\mathrm{v.r.}}/\theta_c$ at $R/4R_c\to\infty$ amounts
$\pi/2$ for positively charged particles, and 1 for negatively
charged particles. More precisely, this is the outer edge of the
angular distribution. The deflection angle mean value involves an
$\mathcal{O}(R_c/R)$ correction:
\begin{equation}\label{theta-mean}
    \left\langle\theta\right\rangle=\int d\theta\theta\frac{dw_{\mathrm{real}}}{d\theta}=\int d\theta\theta\frac{dw_{\mathrm{v.r.}}}{d\theta}=\Bigg\{\begin{array}{c}
                                         \!\!-\frac{\pi}{2}\theta_c\left(1-\frac{2R_c}{R}\right) \qquad\quad \mathrm{for\, pos.\, charged\, particles}\\
                                         \!\!-\theta_c\left(1-1.3\frac{R_c}{R}\ln\frac{R}{R_c}\right)  \quad \mathrm{for\, neg.\, charged\, particles}
                                       \end{array}
\end{equation}
(the coefficient 1.3 for negative particles needs  more precise
evaluation). For protons, prediction (\ref{theta-mean}) was compared
with experiment in \cite{ref:Bond-VR}, see also
Fig.~\ref{fig:sigmaVR} below.

The second moment, mean square deflection, under assumption
(\ref{convol}) possesses the property of additivity, even for
non-gaussian ${dw_{\mathrm{v.r.}}}/{d\theta}$:
\begin{equation}\label{sigma2-add}
    \sigma^2_{\mathrm{real}}\approx\int^\infty_{-\infty}d\theta\left(\theta-\left\langle\theta\right\rangle\right)^2\int^\infty_{-\infty}d\alpha\frac{e^{-\frac{(\theta-\alpha)^2}{2\sigma^2_{\mathrm{am}}}}}{\sqrt{2\pi}\sigma_{\mathrm{am}}}\frac{dw_{\mathrm{v.r.}}}{d\alpha}=\sigma^2_{\mathrm{am}}+\bar\sigma^2_{\mathrm{v.r.}},
\end{equation}
where
$\bar\sigma^2_{\mathrm{v.r.}}=\int^\infty_{-\infty}d\theta(\theta-\left\langle\theta\right\rangle)^2\frac{dw_{\mathrm{v.r.}}}{d\theta}$.
For positively charged particles, acquiring rectangular intrinsic
final angular distribution at $\frac{4R_c}{R}<1$, evaluation of
$\bar\sigma_{\mathrm{v.r.}}$ gives
\begin{equation}\label{model-sigma}
    \bar\sigma_{\mathrm{v.r.}}=\frac{\pi}{\sqrt3}\frac{R_c}{R}\theta_c\qquad\quad
\left(\mathrm{for\, positively\, charged\, particles}\right).
\end{equation}
Experiment \cite{ref:Scandale-Rdependence} used procedure
(\ref{sigma2-add}), i.e., formula
$\bar\sigma_{\mathrm{v.r.}}=\sqrt{\sigma^2_{\mathrm{real}}-\sigma^2_{\mathrm{am}}}$,
to determine $\bar\sigma_{\mathrm{v.r.}}$. Comparison of
experimental data with Eq.~(\ref{model-sigma}) is shown in
Fig.~\ref{fig:sigmaVR} by dashed line.

\begin{figure}[h]
\includegraphics[width=15pc]{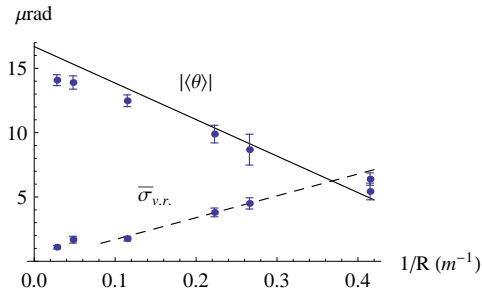}
\caption{\label{fig:sigmaVR} Deflection angle mean value
(Eq.~(\ref{theta-mean}), solid line) and rms deviation
(Eq.~(\ref{model-sigma}), dashed line) for a parabolic model of
inter-planar potential compared with CERN SPS experimental data
\cite{ref:Scandale-Rdependence} (protons, $E$=400 GeV).}
\end{figure}

For negatively charged particles, where the final profile is skew,
its third moment is also non-zero: $\int
d\theta\left(\theta-\left\langle\theta\right\rangle\right)^3\frac{dw_{\mathrm{real}}}{d\theta}\approx\int
d\theta\left(\theta-\left\langle\theta\right\rangle\right)^3\frac{dw_{\mathrm{v.r.}}}{d\theta}\neq0$.

\subsection{Conditions for quality VR deflection} Joint
consideration of beam spread in the continuous potential, multiple
scattering in the silicon material, and the initial beam angular
divergence has led \cite{ref:Bond-VR} to a system of conditions
needed for practical application of VR to beam steering at
accelerators:
\begin{equation}\label{condition-system}
    \frac{L}{1\,\mathrm{mm}}<\frac E{100\,\mathrm{GeV}}<\frac
R{\mathrm{m}},\,\left(\frac{20\,\mu\mathrm{rad}}{\sigma_0}\right)^2,
\end{equation}
where $\sigma_0$ is the initial beam rms divergence. Conditions
(\ref{condition-system}) are fairly well met in experiments
\cite{ref:Scandale-Rdependence}.

\subsection{VR at $\theta_0\sim\theta_c$}
The theoretical description of VR in
\cite{ref:Maisheev-VR,ref:Bond-VR} focussed on the case
$\theta_0\gg\theta_c$, when the process may be regarded as
independent of the crystal boundaries. But in experiments, VR is
often investigated in parallel with channeling, with a continuous
passage from one regime to another. In the transition regime, the
final beam angular distribution must acquire substantial
$\theta_0$-dependence. Investigation of this dependence is important
as well (see \cite{ref:Dabagov-VR}).


\section{Radiation in thin uniformly bent crystals}

Observation of inelastic processes evoked by a fast particle
traveling a bent crystal may be used for extraction of information
about particle trajectories at real conditions. Among possible
inelastic processes, the coherent radiation in the near-forward
direction\footnote{The direct radiation from the fast passing
particle is only significant for electrons and positrons, whereas
for beam deflection applications the primary interest is on protons.
However, ultra-relativistic positrons have essentially the same
trajectory as protons of equal energy, so one indirectly can test
proton passage through experiments with positrons. } benefits from
existence of a local correspondence between the photon $\omega$ and
the longitudinal coordinate $z$ it is emitted from. This relation is
particularly simple (geometric rather than dynamic) at large
$\omega$, corresponding to $z$ away from the VR area. There, the
particle deflection from the straight line is only perturbative,
opposite in sign for electrons and positrons. Here we will confine
ourselves to discussion of the latter simplest case -- coherent
bremsstrahlung in a bent crystal (CBBC).


\subsection{CBBC intensity as a sum of infinitesimal straight-crystal contributions}
The first, analytic, calculation of coherent bremsstrahlung in a
bent crystal was undertaken in \cite{ref:Arutyunov} (in appendix of
a paper primarily dedicated to \emph{channeling} radiation in a
uniformly bent crystal). Fourier decomposition of the bent crystal
continuous potential yielded Fresnel functions, which describe both
volume effects (step-like asymptotics), and the oscillations due to
edge effects. But at practice usually the edge effects are
negligible, and the Fresnel functions may be replaced by a Heavyside
unit-step function, which led to Eq. (A.16) of \cite{ref:Arutyunov}:
\begin{equation}\label{A16Arut}
    d\sigma^B(\omega,\theta_0)=\frac1{\Delta\psi(R,L)-\theta_{\min}}\int^{\Delta\psi(R,L)}_{\theta_{\min}}d\sigma^S(\omega,\theta_x;L)d\theta_x,
\end{equation}
with $d\sigma^S$, $d\sigma^B$ the radiation differential
cross-sections in a straight and in a bent crystal, $\Delta\psi$ the
angle between the particle velocity and active crystallographic
planes on the crystal edge, and
$\theta_{\min}\simeq\frac{q_{\min}d}{2\pi}\ll\Delta\psi$. This
representation in form of unweighted $\theta_x$-averaging is only
valid at $R=\mathrm{const}$. Eq.~(\ref{A16Arut}) also assumes
symmetric orientation of the beam with respect to the particle
trajectory (when $\Delta\psi$ at the entrance from the crystal
equals $\Delta\psi$ at the exit), though generalization to an
asymmetric case is straightforward. In fact, at practice such a
generalization may be necessary even if the crystal is oriented
symmetrically with respect to the beam axis, since non-negligible
beam divergence compared to the crystal half-bending-angle makes the
orientation asymmetric for individual particles
\cite{ref:Bond-JPhys}.

A generic representation (valid at variable $R$ and arbitrary
orientation), which can be obtained based on the stationary phase
approximation \cite{ref:Bond-CBBC} \footnote{The replacement of
Fresnel functions by step functions in \cite{ref:Arutyunov} is
equivalent to application of the stationary phase approximation in
the original Fourier integrals.}, reads
\begin{equation}\label{dECBBC-dz}
    \frac{dE_{\mathrm{CBBC}}}{d\omega}=\int^{L/2}_{-L/2}dz\frac{dE_{\mathrm{straight}}}{dzd\omega}\Big|_{\theta_x=|\xi'(z)-\theta_0|},
\end{equation}
where $\xi(z)$ is the transverse coordinate of any of the bent
atomic planes. If one utilizes here the known formula for the
spectrum of coherent bremsstrahlung in a straight crystal
\begin{equation}\label{dECBBC-straight}
    \frac{dE_{\mathrm{straight}}}{dzd\omega}=\frac{e^2F^2_{1}d^2}{2\pi^4m^2\theta_x^2}\frac{E'^2}{E^2}q_{\min}
    \sum^\infty_{n=1}\Theta\!\left(\!n\!-\!\frac{q_{\min}d}{2\pi|\theta_x|}\!\right)\!\frac{c^2_n}{n^{4+2\epsilon}}\!\left(\!1\!-\!\frac{q_{\min}d}{n\pi|\theta_x|}\!+\!\frac{q^2_{\min}d^2}{2n^2\pi^2\theta_x^2}\!+\!\frac{\omega^2}{2EE'}\!\right)\!,
\end{equation}
with $ q_{\min}(\omega)=\frac{m^2\omega}{2EE'}$, $E'=E-\omega$,
$\theta_x(z)=\xi'(z)-\theta_0$, equation (\ref{dECBBC-dz}) turns to
explicit Eq.~(30) of paper \cite{ref:Bond-CBBC}.
To see the correspondence of Eqs. (\ref{A16Arut}) and
(\ref{dECBBC-dz}), one observes that in a uniformly bent crystal
$\xi(z)=\frac{z^2}{2R}+\mathrm{const}$, so integration variables
$\theta_x=|z/R-\theta_0|$ and $z$ are linearly related. In the ratio
of $d\sigma^{S}\propto L$ and $\Delta\psi\propto\frac{L}{R}$ the
crystal thickness $L$ cancels out, so $\frac{dE}{d\omega}\propto R$.
From the viewpoint of Eqs.~(\ref{dECBBC-dz}, \ref{dECBBC-straight}),
the proportionality of the radiation spectrum to the crystal bending
radius arises as
\begin{equation}\label{eqqq3}
    \frac{dE_{\mathrm{CBBC}}}{d\omega}\propto \frac{dz}{dq}\sim
Rd\sim l^2_{\mathrm{EFC}};
\end{equation}
see also Eq.~(\ref{eee}) below.

For a uniformly bent crystal, further on, it is straightforward to
accomplish integration over $\theta_x$ and arrive at Eq.~(31) of
\cite{ref:Bond-CBBC}, presenting the spectrum through a cubic
polynomial function $D$. Averaging over angular distribution in the
initial beam and account of multiple scattering in the crystal
volume is also feasible analytically.

\subsection{Coherence lengths and typical photon energies}

Evaluation of the spectrum of bremsstrahlung produced under a
definite force $F(t)$ action on a charged particle often proceeds in
two steps: (i) Fourier-decomposition of the external force, i.e.,
distinguishing virtual photons absorbed by the particle; (ii)
conversion of the force spectrum to the radiation spectrum. An
important physical scale characterizing each stage is its coherence
length. In general, real photon emission ($l_{\mathrm{form}}$) and
virtual photon absorption ($l_{\mathrm{EFC}}$) coherence lengths
need not be related. Indeed, in simplest cases one has
$l_{\mathrm{EFC}}\to0$ (stochastic scattering in an amorphous
medium), or $l_{\mathrm{EFC}}\to\infty$ (uniform external field),
but $l_{\mathrm{form}}=\frac{2EE'}{m^2\omega}$ always stays finite,
providing for those cases the only coherence length to refer to.
From the experience of the mentioned simplest cases, it had become
common to call $l_{\mathrm{form}}$ simply `coherence length',
without specification of the nature of the coherence.

On the other hand, if the bremsstrahlung problem may be treated in
the dipole approximation, the photon emission coherence is strictly
related with the virtual photon absorption coherence: it reduces to
Lorentz-rescaling of the frequency and integration of the intensity
over typical $\sim\gamma^{-1}$ emission angles (cf.
Eq.~(\ref{omega0}) below). Thence, it suffices to analyze the
coherence in the external field Fourier decomposition.

In crystals, $l_{\mathrm{EFC}}$ acquires a finite size. In a
straight crystal, this is just the distance between the planes
crossed along the particle path; the irradiation resonance condition
requires $l_{\mathrm{form}}$ to have the same value. Then,
\begin{equation}\label{lcoh-straight}
    l_{\mathrm{form}}\sim l_{\mathrm{EFC}}\simeq{d}/{\theta_0}\qquad\qquad\qquad (\mathrm{straight\,
crystal}).
\end{equation}
At that, the typical radiation energy
\begin{equation}\label{omega0}
    \omega_0\sim 2\gamma^2 l^{-1}_{\mathrm{form}}\sim
\gamma^2\theta_0/d \qquad\qquad\qquad (\mathrm{straight\, crystal})
\end{equation}
is just the Lorentz-rescaled plane-crossing frequency. The radiation
spectral intensity is estimated as (cf. (\ref{dECBBC-straight}))
\begin{equation}\label{straight-spectrum}
    \qquad\frac{dE}{d\omega}=\frac{dE}{dzd\omega}L\propto \left(\frac{eF}{m}\right)^2\frac{d^2}{\theta_0^2}\frac{L}{l_{\mathrm{form}}(\omega)},\quad l_{\mathrm{form}}^{-1}=q_{\min}(\omega)\lesssim\theta_0/d\qquad (\mathrm{straight\,
crystal}).
\end{equation}

Less trivial situation emerges in a bent crystal. There, the local
frequency of plane crossing varies along the crystal, whereby there
are 2 spatial scales:
\begin{equation}\label{lEFC}
    l_{\mathrm{EFC}}=\sqrt{2Rd}\qquad\qquad (\mathrm{bent\,
crystal}),
\end{equation}
on which the integrals from oscillatory gaussians ($\int dz
e^{i\frac{z^2}{2Rd}}...$) converge, and the longitudinal geometrical
scale of the crystal (say, its thickness $L$) determining the
limiting frequency of plane crossing. If a tangency point of the
particle trajectory with the family of bent planes occurs within the
crystal, then
\begin{equation}\label{eee}
    \frac{dE}{d\omega}\propto \left(\frac{eF}{m}\right)^2
l^2_{\mathrm{EFC}} \qquad\qquad (\mathrm{bent\, crystal}).
\end{equation}

Comparing Eqs.~(\ref{eee}) and (\ref{straight-spectrum}), one sees
that in both cases the radiation intensity is proportional to
$l^2_{\mathrm{EFC}}$, which is in line with the understanding
\cite{ref:Ryazanov} of coherence length as the length at which
radiation amplitudes add up. Besides that, (\ref{straight-spectrum})
involves a factor $\frac{L}{l_{\mathrm{form}}}$ arising due to
spectral overlap of radiation generated within the crystal thickness
in different coherence intervals (pile-up factor). In a bent
crystal, where radiation from different coherence intervals does not
overlap, factor $\frac{L}{l_{\mathrm{EFC}}}$ enters the expression
for the spectrum extent
\begin{equation}\label{omega-end}
    \omega_{\mathrm{end}}\propto\min\left\{4\pi\gamma^2{L}/{l^2_{\mathrm{EFC}}}, E\right\}
\end{equation}
instead of intensity (\ref{eee}).

Since $l_{\mathrm{EFC}}$ and $l_{\mathrm{form}}$ appear to be
important physical quantities, it is worth estimating their typical
values. For $R\sim 20$ m, $d\approx 2$ ${\AA}$, Eq.~(\ref{lEFC})
gives $l_{\mathrm{EFC}}\sim 100$ $\mu\mathrm{m}$, whereas ratio
$\frac{l_{\mathrm{form}}}{l_{\mathrm{EFC}}}\gtrsim
\frac{l_{\mathrm{EFC}}}L$ is usually small (see Eq.~(\ref{lEFCllL})
below). Therefore, without breaking the picture of CBBC, the crystal
thickness may be decreased down to fractions of millimeter. Lastly,
concerning the multiple scattering influence on the radiation, it is
quantified by parameter
$\frac{l_{\mathrm{EFC}}}{l_{\mathrm{mult}}}$, which is larger than
LPM parameter $\frac{l_{\mathrm{form}}}{l_{\mathrm{mult}}}$, and
therefore is more critical.

\subsection{Locality of CBBC generation vs. straight crystal limit}
The observation that representation (\ref{A16Arut}) contains a
straight crystal limit may appear surprising from the viewpoint that
local CBBC theory is based on the stationary phase approximation,
whose condition assumes
\begin{equation}\label{lEFCllL}
    l_{\mathrm{EFC}}\ll L \qquad\qquad (\mathrm{CBBC \,\, locality\,
condition}).
\end{equation}
As $R$, and therewith $l_{\mathrm{EFC}}$, increases, condition
(\ref{lEFCllL}) must break down. Nevertheless, this does not destroy
the convergence of the integral, provided
\begin{equation}\label{eqqq17}
    {4d}/{|\theta_0|}\ll L\qquad\qquad (\mathrm{condition\, of\, many\text{-}interval\, crossing\, in\, a \, staight\,
crystal}),
\end{equation}
i. e., the particle crosses a large number of inter-planar intervals
even in a straight crystal. If (\ref{eqqq17}) holds, with the
increase of $R$ one sooner arrives at a condition
\begin{equation}\label{cond}
    |\theta_0|\gg{L}/{2R}
\end{equation}
than at (\ref{lEFCllL}). That implies that variation of the local
plane-crossing frequency in the bent crystal is smaller than the
frequency mean value, which is equivalent to near straightness of
the crystal, regardless of whether condition (\ref{lEFCllL}) holds
or not. In \cite{ref:Bond-CBBC} it was demonstrated that under
condition (\ref{cond}), Eqs.~(\ref{dECBBC-dz}-\ref{dECBBC-straight})
indeed turn to the familiar formula of CBBC in a straight crystal,
with the crystal bending radius dropping out. Hence, further
increase of the radius will be inconsequential. The
\emph{linear}-exponent oscillatory integral convergence will be
achieved at length (\ref{lcoh-straight}) within the crystal
thickness without the aid of crystal curvature. Together, conditions
(\ref{lEFCllL}-\ref{eqqq17}) may be cast into a universal expression
for the convergence length: $
    l_{\mathrm{conv}}=\min\left\{l_{\mathrm{EFC}}, {4d}/{|\theta_0|}\right\}.
$

\subsection{Radiation polarization}

For planar orientation of the crystal, when the particle is mainly
subject to the force orthogonal to the active planes, there must be
a significant net polarization of the cone of emitted
bremsstrahlung, directed orthogonally to the planes. According to
\cite{ref:Bond-Polarization}, the polarization degree equals
\begin{equation}\label{Paggr}
\mathtt{P}_{\mathrm{net}}\left({\omega}/E\right)=\frac{N^2}2\frac1{1+\frac{3\omega^2}{4EE'}},
\end{equation}
where the azimuthal anisotropy parameter $N$ for planar orientation
must be close to 1. Thereat, the maximal value of
$\mathtt{P}_{\mathrm{net}}$ is 50\% at $\omega\ll E$, while at
$\omega\to E$, $E'\to 0$, $\mathtt{P}_{\mathrm{net}}\to0$.

\section{Conclusions}
The problem of particle interaction with bent crystals in planar
orientation contains many opportunities for analytic description.
Some successes have already been achieved, linear laws of
Fig.~\ref{fig:sigmaVR} being an example. Ultimately, for this class
of problems there is a good perspective to calculate all the
relevant observables without MC simulation, although the number of
problems to deal with is large, and the goal remains remote
presently.






\acknowledgments

The author wishes to Organizers of Channeling-2010 conference for
local support. My special gratitude is to Prof. N.F.~Shul'ga who
introduced me to the subject of particle passage through matter when
I was a student, and to Dr. A.V.~Shchagin, who helped me keep in
course of new experimental developments in this area.

\end{document}